\documentclass[11pt,a4paper]{article}
\usepackage[hyperref]{acl2017}
\usepackage{times}
\usepackage{latexsym}
\usepackage{graphicx}
\usepackage{url}
\usepackage{amsmath}
\usepackage{graphicx}
\usepackage{caption}
\usepackage{subcaption}

\aclfinalcopy 

\title{DeepSeek: Content Based Image Search \& Retrieval}

\author{Tanya Piplani\\
 School of Information\\
 UC Berkeley \\
 {\tt tanyapiplani@berkeley.edu}}
 
\date{}

\begin{document}
\maketitle
\begin{abstract}
Most of the internet today is composed of digital media that includes videos and images. With pixels becoming the currency in which most transactions happen on the internet, it is becoming increasingly important to have a way of browsing through this ocean of information with relative ease. YouTube has 400 hours of video uploaded every minute and many million images are browsed on Instagram, Facebook, etc. Inspired by recent advances in the field of deep learning and success that it has gained on various problems like image captioning ~\cite{karpathy2015deep} and \cite{DBLP:journals/corr/XuBKCCSZB15}, machine translation ~\cite{DBLP:journals/corr/BahdanauCB14}, word2vec , skip thoughts \cite{DBLP:journals/corr/KirosZSZTUF15}, etc, we present DeepSeek a natural language processing based deep learning model that allows users to enter a description of the kind of images that they want to search, and in response the system retrieves all the images that semantically and contextually relate to the query. Two approaches are described in the following sections. 
\end{abstract}

\section{Introduction}

Image search is a very challenging problem which is subject of active research today. All major players like Amazon, Google, Apple, etc provide a solution for the same. However all of these have limitations. For instance, Amazon's image search uses computer vision to retrieve similar images. While accurate in most cases, the biggest issue here is that the user needs to input a image based query, which might most of the times be not easily available. Apple in its devices provides option to search for images through small phrases like "food", "birthday", etc. Because of being limited by the amount of tokens that can be accurately processed, the expressivity is severely limited. Also this search is of course limited to the number of images on a device. Some other solutions like Google's image search use meta-data which may be quite mis-leading. 

To overcome all of the problems we propose an end-to-end way of image search and retrieval through text based queries by using natural language processing. In the next section we describe the data that we will be using and the approaches. 

\section{Related Work}
A lot of work has been done in the field of content based image retrieval. \cite{DBLP:journals/corr/ZhouLT17} Specially, two pioneering
works have paved the way to the significant advance in
content-based visual retrieval on large-scale multimedia
database. The first one is the introduction of invariant local
visual feature SIFT \cite{Lowe:2004:DIF:993451.996342}. The second work is the introduction of
the Bag-of-Visual-Words (BoW) model \cite{1238663}. Leveraged from
information retrieval, the BoW model makes a compact
representation of images based on the quantization of the
contained local features and is readily adapted to the classic
inverted file indexing structure for scalable image retrieval. Image representation originates from the fact that the
intrinsic problem in content-based visual retrieval is image
comparison. For convenience of comparison, an image is
transformed to some kind of feature space. The motivation
is to achieve an implicit alignment so as to eliminate the
impact of background and potential transformations or
changes while keeping the intrinsic visual content distinguishable. Traditionally, visual features are heuristically designed and
can be categorized into local features and global features.

\begin{figure*}[h]
\begin{center} \includegraphics[width=\linewidth]{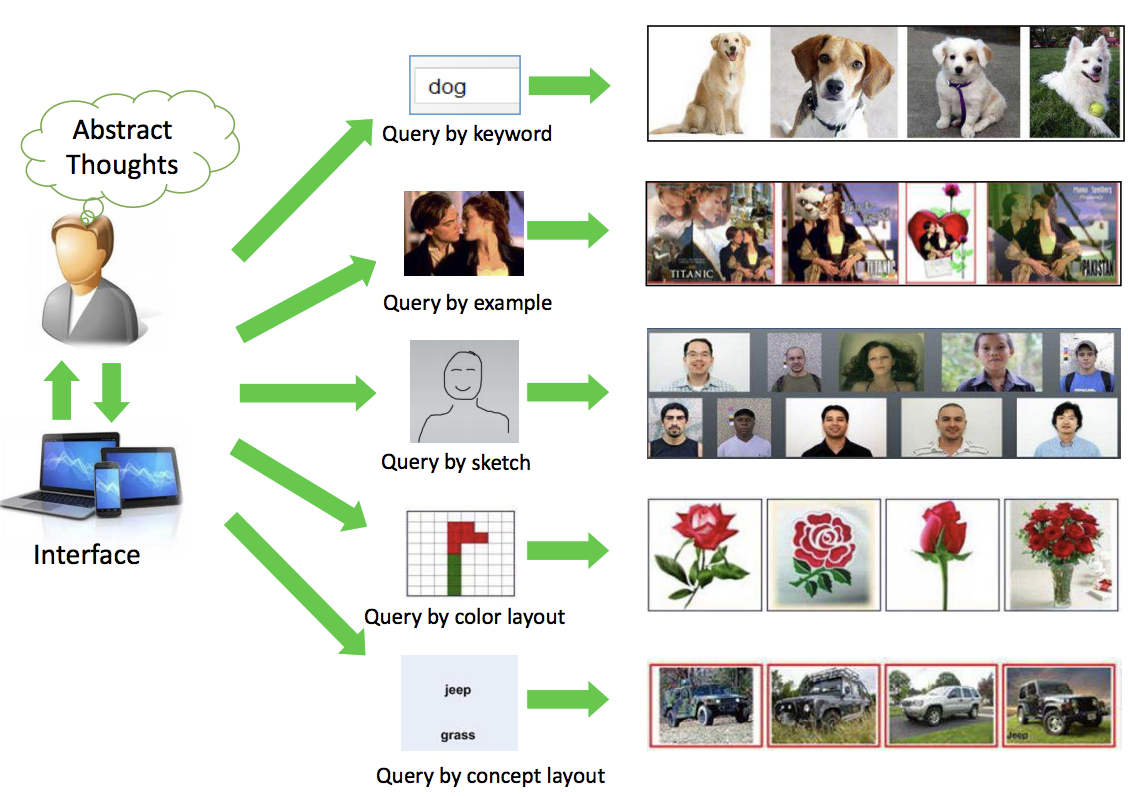}
\end{center}
\caption{ Illustration of different query schemes with the corresponding retrieval results.}
\end{figure*}

Besides those hand-crafted features, recent years have witnessed
the development of learning-based features. The biggest problem with these systems are that the query formulation is in terms of an input image, that is transformed into a feature representation; based on which, the images in the dataset are ranked. This is fundamentally flawed as it requires the user to have a similar image to begin with. Some approaches are also outlined that take text as input for searching images on the web, like Google. But most of the emphasis here is actually doing text retrieval \cite{6739088} and returning images that are associated with the retrieved text. This is by definition a weak learning approach, where the problem is being solved indirectly.

Thus all of the above systems are lacking in there effort to provide an efficient solution for text based semantic image retrieval. Hence, we propose a method to semantically embed the text and the image in the same space, so that we can more efficiently and accurately retrieve images based on a text query.

\section{Our Method}

In this section we describe two approaches that we want to try, to solve the problem detailed above.

\subsection{Caption Based Retrieval}
\begin{figure*}[h]
\begin{center} \includegraphics[width=\linewidth]{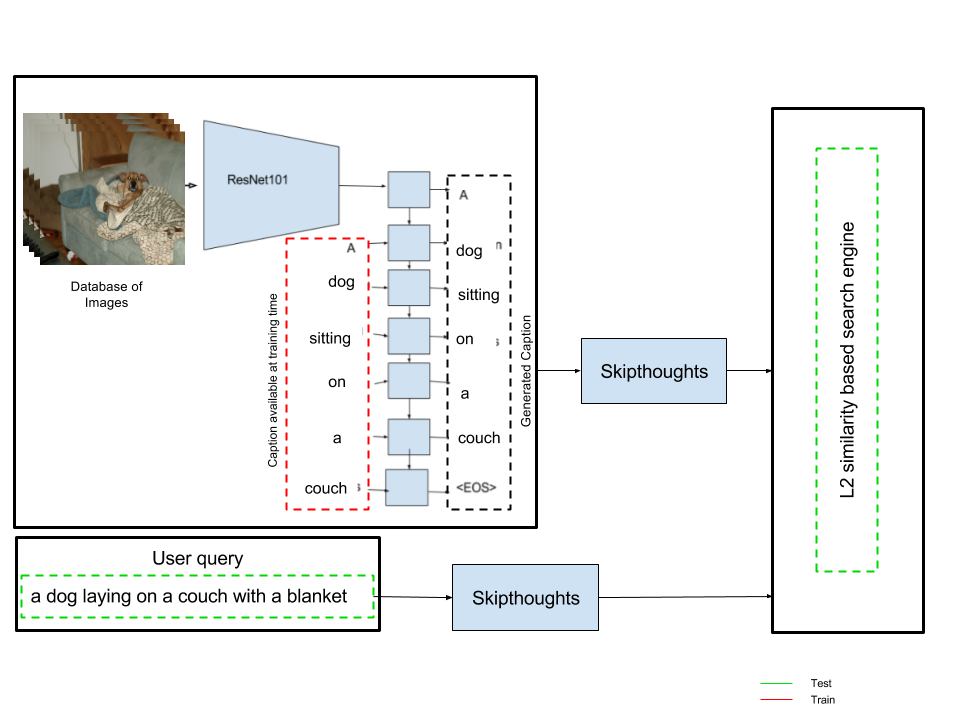}
\end{center}
\caption{The end-to-end retrieval system for the caption based image retrieval pipeline.}
\end{figure*}

A lot of work has been done in the field of Image Caption Generation. The problem of Image-Caption Generation deals with generating a single phrase caption describing the scene in the image. A state-of-the-art convolutional neural network architecture is used for extracting image based features and converting the input into a vector of embedding that is semantically rich. This vector can then be used to train different task like classification, detection or can be used as part of a pipeline for some other task. For our purpose we will use this for initializing a language model. The language model is a Long Short Term Memory based architecture that tries to model $P(S_t \vert h_{t-1}, x_t, S_{t-1})$ where $S_t$ is the word at time $t$, $h_{t-1}$ is the hidden state of the LSTM at time $t-1$ and $x_t$ is the input to the LSTM cell at time $t$. 
At each time step, a softmax is used to output the probability of all words in the vocabulary. 
\begin{figure*}[h]
\begin{center} \includegraphics[width=\linewidth]{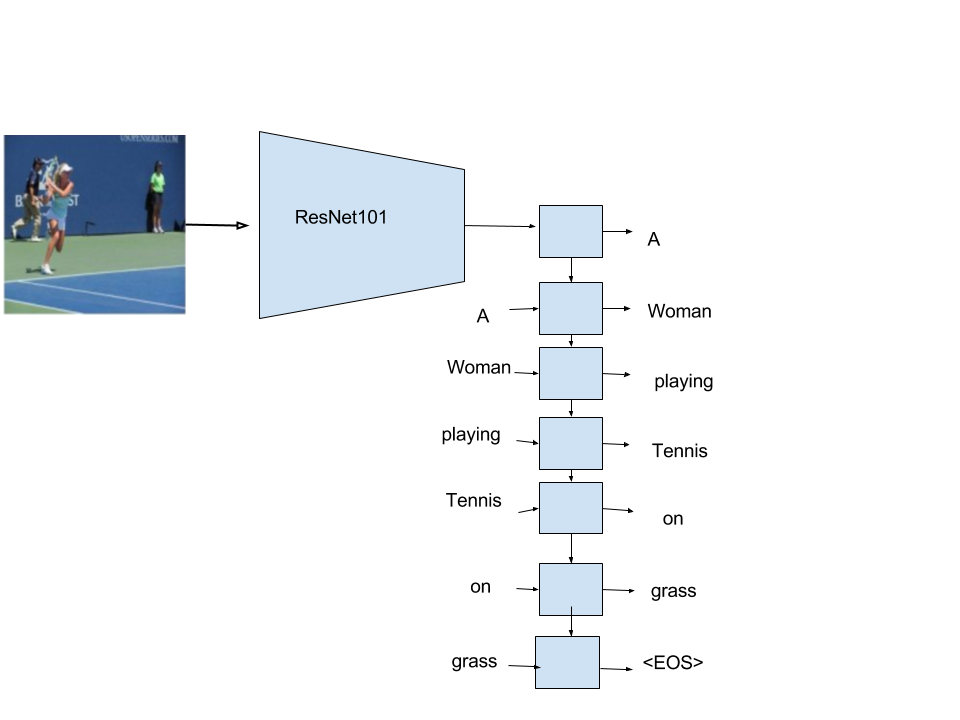}
\end{center}
\caption{Caption Generation}
\end{figure*}

For our experiments we used ResNet-101 \cite{DBLP:journals/corr/HeZRS15} as the feature extraction backbone. We initialized the network with the weights pre-trained from a MS-COCO object detection \cite{DBLP:journals/corr/LinMBHPRDZ14} task and then finetuned it for the task of caption generation on the MS-COCO dataset. Once the captions are generated, we used the skip thought \cite{DBLP:journals/corr/KirosZSZTUF15} model that converts the captions into a vector embedding. Once the captions are converted into vectors, we do the same to the query that is provided by the user. Then the retrieval of images is performed by minimizing the L2 distance between the two vectors ( that of the query and the caption associated with the image). 

\subsection{Embedding Space Retrieval}
\begin{figure*}[h]
\begin{center} \includegraphics[width=\linewidth]{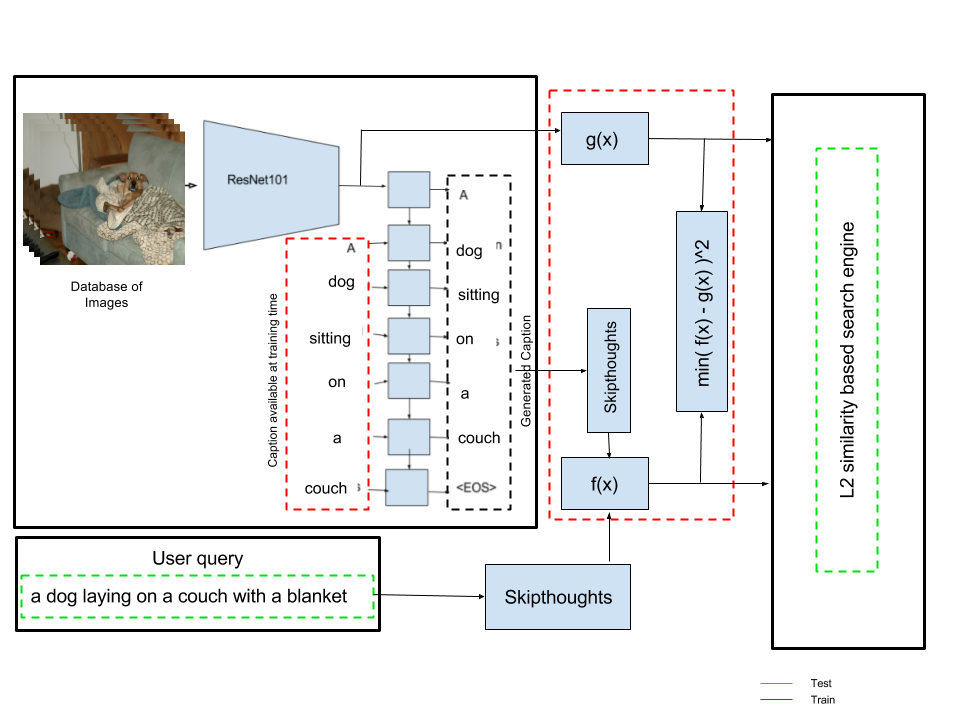}
\end{center}
\caption{The end-to-end retrieval system for embedding space image retrieval}
\end{figure*}

This model consists of the following components:-
\begin{enumerate}
    \item 
    A pre-initialized state-of-the-art convolutional neural network (ResNet-101) is used to extract semantic information from the image frames to construct features that represent the content of the image. We call this vector $V(x_i)$.\\
    \begin{align}
        V(x_i) &= CNN(x_{i})
    \end{align}
    \\where $i\in \{1, \dots N\}$ is one of the images in the dataset of $N$ images.
    \item
    
    The captions that are related to this image are also projected into a semantic feature representation space using the skipthoughts model.
    
    \begin{align}
        U(c^{k}_i) &= \Gamma(c_{i})
    \end{align} 
    
    where $c_i$ is the $k-th$ caption related to the image and $\Gamma$ is the skipthoughts model.
    
    \item

      A projection is then applied to both of these features to create an embedding space which can be learned by minimizing the L2 distance of these vectors.
      
    \begin{align}
        E_v(V(x_i)) &= W_v V(x_i) + b_v
    \end{align}  
      
      and
    \begin{align}
        E_u(W(c^k_i)) &= W_u U(c^k_i) + b_u
    \end{align}  
    \\ where both $E_v(V(x_i))$ and $E_u(W(c^k_i))$ $\in \mathbf{R}^d$.
      \item
      
      The objective function is defined as
     :-\\
    \begin{align}
        \mathbf{L}(E_u(U(c^k_i)),E_v(V(x_i)) )  &=\\
        \vert E_U(U(c^k_i)) - E_v(V(x_i))\vert ^2
    \end{align}  
      \\Thus we end up with a space where both the image and its related captions are close to each other. This space can then be used to project the query from the user, and retrieve the images based on their L2 distances. 

\end{enumerate}

\section{Experiments}
\subsection{Caption Generation}

For our first method, we first train a caption generation model. Here the idea is to convert image into its semantically rich equivalent text representation. The caption thus generated is transformed into a vector using skipthoughts  (explained later). Thus each image is indirectly converted into its semantically rich feature representation. 

\subsubsection{Dataset}
We trained a caption generation model on the MS COCO dataset \cite{DBLP:journals/corr/LinMBHPRDZ14}. This dataset contains images of complex
everyday scenes containing common objects in their natural context. Objects are labeled using per-instance segmentations to aid in
precise object localization. The dataset contains photos of 91 objects types that would be easily recognizable by a 4 year old, with a
total of 2.5 million labeled instances in 328k images. Images have been human annotated with 5 captions each. The 2014 version of this dataset has 80k images for training, 40k for validation and 20k for testing. We follow this same setup. For image retrieval we use the same test set as that for the caption generation, i.e. 20k images of the MS COCO dataset. For the caption based retrieval, there is no training phase except for training the caption generation model itself. For the embedding based retrieval model, the training set of 80k images of the MS COCO dataset are used. Once the vectors are extracted, we define the L2 loss between them as given by equation $5$.

\subsubsection{Setup and Training}

For the caption generation model we use a batch size of 128 with images of size 224x224. Training is done on a Nvidia Titan-X GPU. We use Adam optimizer with momentum and the default settings ( $\beta_1 = 0.99$ and $\beta_2 = 0.9999$). The learning rate is set to 1e-3 and is decayed exponentially. The learning rate for the CNN is set to be an order of magnitude smaller. Gradients for the LSTM are clipped at 10. Training is allowed until convergence. At the time of this report, we were able to run 70k iterations for a CIDEr score of 0.7.

\begin{figure*}[ht]
\begin{center} \includegraphics[width=\linewidth]{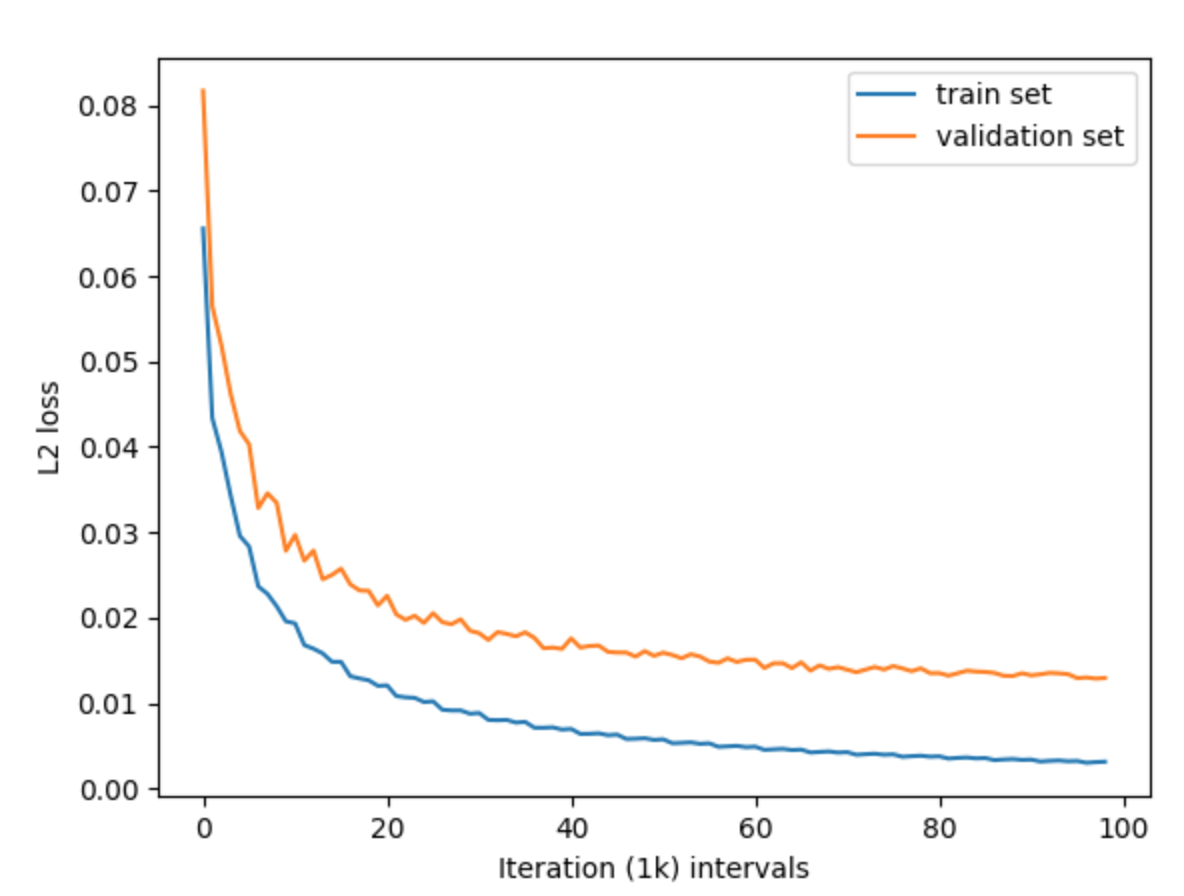}
\end{center}
\caption{The L2 loss curve for the training and validation set for the embedding learning task}
\end{figure*}

For the embedding space learning problem, the setup is similar to above. The loss defined in equation $5$ is minimized using Adam optimizer and a batch size of 128 vectors. The entire training set consists of 80k imaged which is the same as the MS COCO datasets training set for caption generation. The learning rate is defined to be 1e-3 and is exponentially decayed over the course of training. Training is allowed to run untill convergence. We also found clipping the gradient at the norm of $1e3$ to be useful and stabilize the training earlier on.

\subsection{Skipthought Vectors}

Skipthought vectors are an approach for unsupervised learning of a generic, distributed sentence encoder. Using the continuity of text from books, first an encoder-decoder 
model is trained that tries to reconstruct the surrounding sentences of an encoded passage. Sentences that share semantic and syntactic properties are thus mapped to similar vector representations. Next a simple vocabulary expansion
method is introduced to encode words that were not seen as part of training, allowing us to expand our vocabulary to a million words. The end result is an off-the-shelf
encoder that can produce highly generic sentence representations that are robust
and perform well in practice. 

\begin{figure*}[h]
\begin{center} \includegraphics[width=\linewidth]{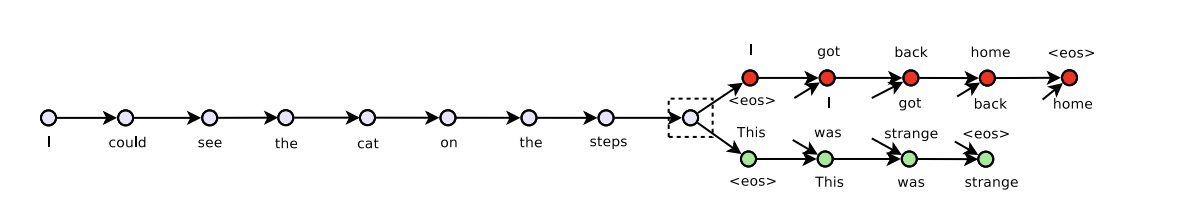}
\end{center}
\caption{The skip-thoughts model. Given a tuple $(s_{i−1}, s_i
, s_{i+1})$ of contiguous sentences, with $s_i$
the i-th sentence of a book, the sentence $s_i$
is encoded and tries to reconstruct the previous sentence
$s_{i−1}$ and next sentence $s_{i+1}$. In this example, the input is the sentence triplet \{ I got back home. I
could see the cat on the steps. This was strange. \} Unattached arrows are connected to the encoder
output. Colors indicate which components share parameters. \{eos\} is the end of sentence token.}
\end{figure*}

For skipthought vector generation, we use the exact same setup as in the original paper. We also do not train skipthought vectors ourselves, and at this time use the model provided on the official github page of the authors. More information about the training of this model can be found in their paper. \cite{DBLP:journals/corr/KirosZSZTUF15}

\subsection{Embedding Spaces for retrieval}
For learning a joint embedding space for both images and text is a difficult task, but has been tackled before as well in several works. We take inspiration from \cite{NIPS2013_5204}, and learn the joint space embeddings by two parallel networks that take the output of the CNN that is part of the caption generation model. These vectors are already very semantically rich but are then embedded to a vector $E_v(V(x_i)) \in \mathbf{R}^d$ in the $d$ dimensional embedding space. The skip thought vectors at training time are taken for the captions that were generated by the caption generation model. The vectors $E_u(W(c^k_i) \in \mathbf{R}^d)$ for the generated caption $c^k_i$ of the image $i$ in our training set is also embedded into the $d$ dimensional space.

\section{Evaluation}

\subsection{Quantitative}

To evaluate different parts of the pipeline, different quantitative metrics are made use of. The caption generation was evaluated using the MS COCO server that makes use of including BLEU, Meteor,
Rouge-L and CIDEr metrics. We compare our model to the existing state of the art system according to the MS COCO caption generation leaderboard.

\begin{table*}[ht]
\centering
\begin{tabular}{|c c c c c|} 
 \hline
 Model & BLUE-1 & METEOR & ROUGE-L & CIDEr-D \\ 
 \hline
 Our Model & 0.928 & 0.320 & 0.693 & 1.092\\ 
 SOTA & 0.953 & 0.375 & 0.734 & 1.270 \\ 
 \hline
\end{tabular}
\caption{Quantitative evaluation of the caption generation model described in this paper as compared to the current state of the art on the MS COCO leaderboard.}
\label{table:1}
\end{table*}

To evaluate our image retrieval systems with each other we come up with the following most relevant metrics. We calculate the precision at three different levels. Precision$@k$ (p$@k$) is defined as follows:-
\begin{align*}
    p@k = \frac{tp@k}{tp@k + fp@k}
\end{align*}

where the true positive event ($tp@k$) is defined when out of the $k$ images that the system retrieves, the correct image is retrieved. And the false positive event $fp@k$ is defined when out of the $k$ images the system is allowed to retrieve none of them are correct. The correct image caption pair comprises of the captions provided in the MS COCO dataset. So given a caption $c_j$ associated with the image $x_j$ in the 20k images in the test set for MS COCO, we perform both the caption based retrieval and the embedding space retrieval as follows. We take the caption $c_j$ associated with the image $x_j$, and embed the caption. All images in the retrieval dataset which is composed of all images in the test set of the MS COCO dataset are also passed through our pipeline to produce their embeddings. If $x_j$ is one of the $k$ images that are retrieved by our system for the query $c_j$, then that is counted as a true positive event $tp@k$ for the calculation of $p@k$. Else it is counted as a false positive event ( $fp@k$).

\begin{table}[h]
\centering
\begin{tabular}{|c c c c c|} 
 \hline
 Model & $p@1$ & $p@3$ & $p@5$ & time(sec) \\ 
 \hline
 ESR & 68.3 & 85.7 & 91.2 & 2.1\\ 
 CBR & 72.9 & 84.9 & 90.5 & 1.7 \\ 
 \hline
\end{tabular}
\caption{Quantitative evaluation of both the approaches for image retrieval described in this paper.}
\label{table:1}
\end{table}

From Table 2 above we see that the $p@1$ for Embedding based retrieval is lower than that of Caption based retrieval and this definitely shows the power of skipthought vectors. While the $p@5$ is more in the case of Embedding based retrieval. This is because the objective function of the model is to get the vector embeddings closer to each other but the first image retrieved might not be the closest and hence the correct image shows up in the top 5 images but may not be the first one retrieved.

Note that the above retrieval operations were not GPU optimized, which would drastically improve the timing for both the approaches. 
\subsection{Qualitative}
In the figure 7 below we compare the output of the two systems i.e. the embedding space retrieval and the caption generation retrieval on the same query to see which one performs better qualitatively. 
\begin{figure*}[h]
        \begin{subfigure}[b]{0.5\textwidth}
                \centering
                \includegraphics[width=.85\linewidth]{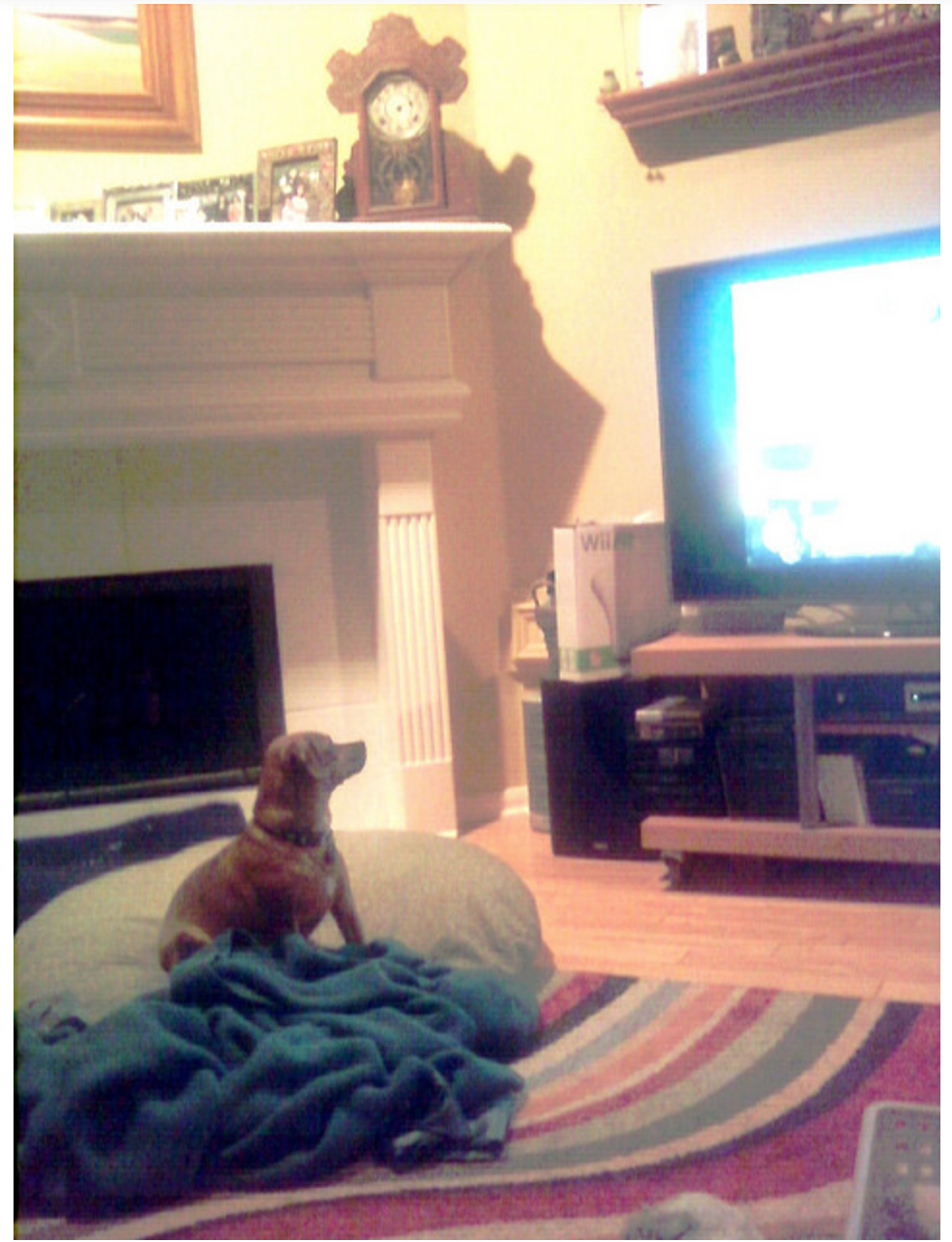}
                \caption{Output of the embedding space retrieval}
                \label{fig:gull}
        \end{subfigure}%
        \begin{subfigure}[b]{0.5\textwidth}
                \centering
                \includegraphics[width=.85\linewidth]{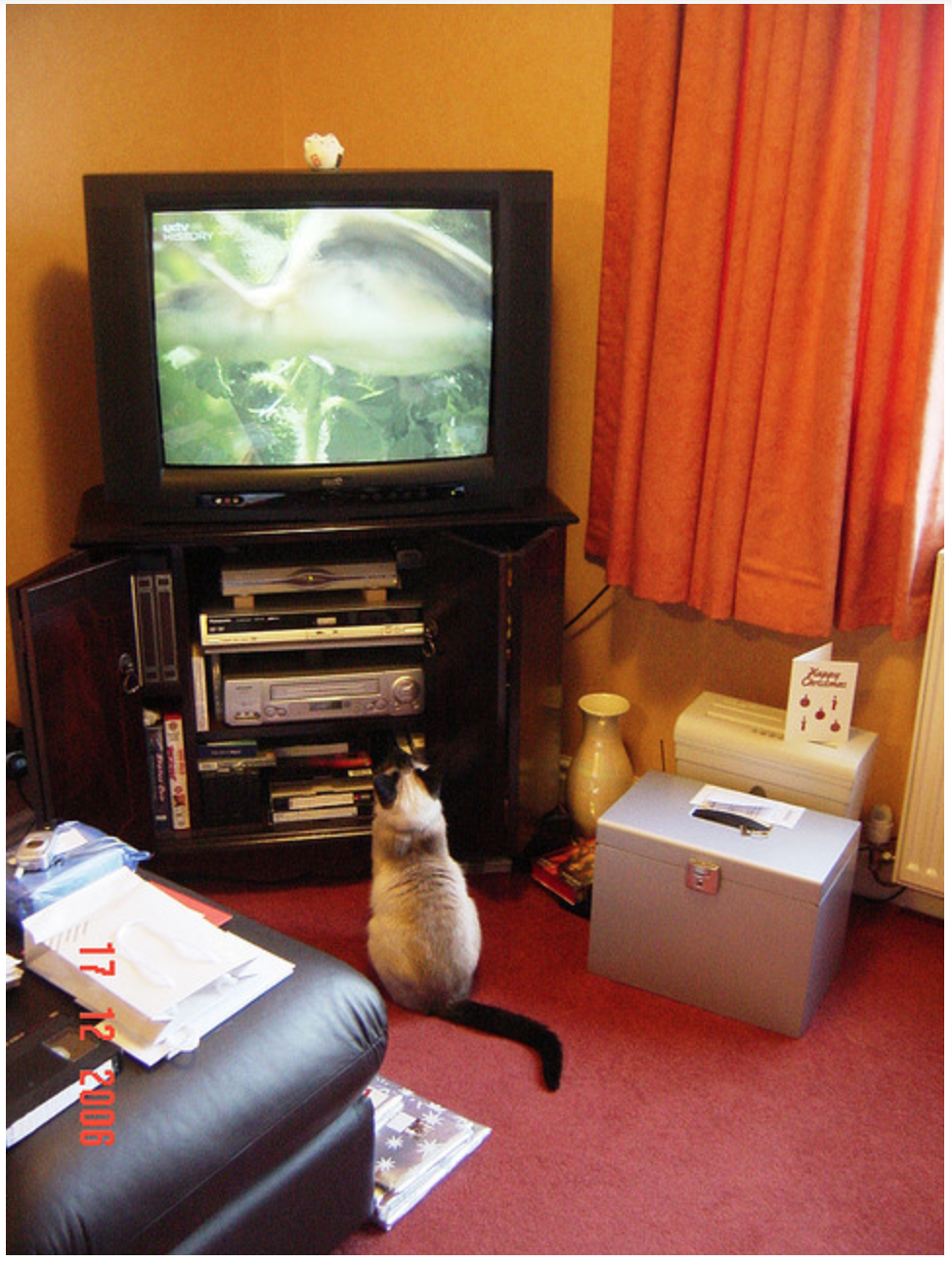}
                \caption{Output of the caption based retrieval}
                \label{fig:gull2}
        \end{subfigure}%
        \caption{Output of both the methods for the query 'A dog watching a television in a room'. It is interesting to see how the exact semantics of not only the objects but the relationship between them are also captured by both of the models. In the results above, we see that not only a animal and a television are present in the scene, but also in both the images these animals are actually watching the television.}\label{fig:animals}
\end{figure*}

\subsection{Conclusions \& Future Work}
As can be seen in the figure $7$, both the methods give nice results qualitatively. It is interesting that not only the correct objects but also the correct semantic relation between various objects in the scene are captured by both the models.

In the future we would like to explore the use of triplet based losses to learn the embedding for Embedding Space Retrieval. This will allow our model not only to be aware about what captions are similar to what images but also which pairs are dissimilar. This would regularize the model quite a lot and also allow more semantically rich features to be learnt. 
Another direction to make the model more user friendly would be to incorporate concepts like knowledge graph, which would allow us to reazon about entities and more. This could allow us to answer queries like "Hector is watching television" instead of "A dog is watching television". We would also like to GPU  optimize the retrieval system to make it much more faster for both the techniques.

\bibliography{acl2017}
\bibliographystyle{acl_natbib}

\end{document}